\documentclass[preprint]{vgtc}                         
\graphicspath{{figures/}{pictures/}{images/}{./}} 
\usepackage{times}                    
         
\usepackage{mathptmx}                  
\usepackage{lipsum}
\usepackage{float}
\usepackage{enumitem}
\usepackage{soul,xcolor}
\usepackage[subpreambles=true]{standalone} %
\usepackage{tikz}
\usepackage{booktabs, tabularx, makecell}
\usepackage[hidelinks,pagebackref,bookmarks]{hyperref}

\definecolor{primaryMainBlock}{HTML}{005c84}
\definecolor{lightMainBlock}{HTML}{DAE9F7}
\definecolor{reviseColor}{HTML}{000000}

\DeclareRobustCommand{\bluecircle}[1]{%
  \tikz[baseline=-.7ex]{%
    \node[
      circle,
      fill=primaryMainBlock,
      inner sep=1pt,
      text=white,
      font=\bfseries\small
    ] (char) {\code{#1}};
  }%
}

\DeclareRobustCommand{\bluebox}[1]{%
  \tikz[baseline=(char.base)]{
    \node[
      draw=primaryMainBlock,         %
      fill=white,        %
      rounded corners=2pt,   %
      inner sep=1.6pt,     %
      outer sep=0pt,
      line width=.8pt   %
    ] (char) {#1};
  }%
}

\DeclareRobustCommand{\ragbox}[1]{%
  \tikz[baseline=(char.base)]{
    \node[
      draw=primaryMainBlock,         %
      fill=lightMainBlock!25,        %
      rounded corners=2pt,   %
      dash pattern=on 2pt off 2pt,
      inner sep=1.6pt,     %
      outer sep=0pt,
      line width=.8pt,   %
    ] (char) {#1};
  }%
}

\newcommand{\orcid}[1]{\href{https://orcid.org/#1}{\includegraphics[width=10pt]{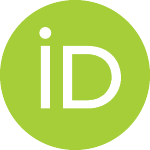}}}

\newcommand{\revise}[1]{\textcolor{reviseColor}{#1}}

\definecolor{inlinequote}{HTML}{FFFBED}
\sethlcolor{inlinequote}
\renewcommand{\copyrightspace}{}

\onlineid{0}
\vgtccategory{Research}
\vgtcinsertpkg

\title{Sycamore: Grounded Synthetic Evaluation for Visualization~Systems}

\author{Huyen N. Nguyen\orcid{0000-0001-6554-2327}\thanks{e-mail: huyen\_nguyen@hms.harvard.edu}\\ %
        \parbox{1.6in}{\scriptsize \centering \vspace{.3em}Harvard Medical School \\[.1em] Los Alamos National Laboratory}
\and Astrid van den Brandt\orcid{0000-0002-3676-1341}\thanks{e-mail: astrid\_vandenbrandt@hms.harvard.edu}\\ %
     \scriptsize Harvard Medical School
\and Nils Gehlenborg\orcid{0000-0003-0327-8297}\thanks{e-mail: nils@hms.harvard.edu}\\ %
     \scriptsize Harvard Medical School}

\abstract{
Evaluating visualization systems in specialized domains, such as genomics, is challenging due to the scarcity of domain experts and difficulty recruiting relevant user types. LLM-based synthetic personas provide an appealing workaround, but they face skepticism as their claims often lack verifiable grounding. We present \emph{Sycamore}, an exploratory approach to evaluating visualization systems with grounded synthetic personas whose responses are restricted to interview evidence. Sycamore enforces provenance with participant citations for each claim and an abstention mechanism when retrieval yields no relevant evidence.
A protocol runner guides the synthetic personas through an evaluation of the live visualization system, producing multi-turn transcripts for qualitative coding. We instantiate the approach on Geranium, a multimodal genomics visualization search engine, with seven evaluators grounded in prior interviews on genomics visualization authoring. We report themes in the resulting transcripts, highlight an emergent requirement absent from the source data, and provide a preliminary comparison to a prior human user study. Recognizing that our findings derive from an exploratory perspective and that synthetic proxies cannot replace human evaluation, we discuss their implications for integrating synthetic evaluators alongside expert user studies. The protocol, codebook, extended results, transcripts, and demo video are in the Supplementary Materials and at \href{https://osf.io/kdfr3/}{\texttt{osf.io/kdfr3}}; the source code and project page are at \href{https://huyen-nguyen.github.io/sycamore/}{\texttt{huyen-nguyen.github.io/sycamore}}.

}

\keywords{Synthetic persona, visualization evaluation, user study, Geranium visualization retrieval, genomics visualization.}

\makeatletter
\let\sy@startsection\@startsection
\renewcommand\@startsection[6]{%
  \sy@startsection{#1}{#2}{#3}%
    {\glueexpr#4*70/100\relax}%
    {\glueexpr#5*80/100\relax}%
    {#6}}
\makeatother

\begin{document}

\maketitle

\vspace{.4em}
\section{Introduction}
\label{sec:intro}

Visualization research has long employed user studies to understand tasks and evaluate systems. In specialized domains such as genomics, however, domain experts are scarce and recruitment is slow. Even when recruitment succeeds, studies may capture only a narrow slice of the target audience. For example, the Geranium user study~\cite{nguyen2026geranium} involved seven bioinformatics researchers; the Blended Interface study~\cite{lyi2024blace} involved twelve domain experts; and the interview study by van den Brandt et al.~\cite{van2024understanding} involved twenty participants despite using multiple recruitment channels.

In parallel, personas based on large language models (LLMs) are increasingly used in design and evaluation, though their adoption in Human-Computer Interaction (HCI) faces skepticism~\cite{myheart2026,deus2024}. The community cautions that synthetic substitutes cannot replace real human users~\cite{ixdf_ai_personas}. This growing concern stems in part from a lack of reliable provenance, where synthetic personas may answer confidently without verifiable sources. One countermeasure to this is \emph{grounding}, which restricts persona outputs to empirical evidence from real users. Grounding operates on a spectrum: while simply assigning roles in prompts offers superficial role-playing, retrieving evidence from coded transcripts provides greater rigor~\cite{hosseini2025llms,truss2026personacite}.

Motivated by this rigorous standard for grounding, we present \textbf{Sycamore} (\underline{Sy}nthetic \underline{C}h\underline{a}racterization for Evaluating Geno\underline{m}ics Visualizati\underline{o}n \underline{Re}trieval), an exploratory framework that assembles synthetic evaluators at query time using persona profiles and coded interview excerpts. To enforce verifiability, evaluators must cite source participants for evidence-based claims and abstain when no retrieved evidence meets the relevance threshold. We instantiate Sycamore on \mbox{Geranium}~\cite{nguyen2026geranium}, a genomics visualization search engine supporting three query \emph{modalities}: text, image, and specification. Evaluators are grounded in profiles of genomics experts and interview excerpts from van den Brandt et al.~\cite{van2024understanding}. This setup allows us to examine how evaluators query the system and respond to retrieval results, identify themes in their responses, and compare these themes with findings from a prior expert study of Geranium. As synthetic evaluators follow a scripted protocol instead of interacting directly with the interface as human users do, our comparison with the prior human study provides only a preliminary assessment of alignment. We also include an ungrounded run to contextualize the grounded responses. Our contributions are as follows:

\begin{itemize}[leftmargin=1.3em,itemsep=.1em]
    \item \textbf{Grounded synthetic evaluators} for visualization retrieval systems, using published expert evidence, citations for each claim, an abstention mechanism for insufficient evidence.
    \item \textbf{An instantiation} on a genomics visualization retrieval system, demonstrating the pipeline and its design decisions.
    \item \textbf{An exploratory analysis} of the resulting transcripts, identifying themes and a preliminary comparison to a prior expert study.
\end{itemize}

\vspace{-.2em}
\section{\revise{Background and} Related Work}
\label{sec:related}

Evaluation in visualization should be grounded in its context of intended use~\cite{grounded_eval, nested}, but in highly specialized domains, it is challenging to recruit enough suitable participants. A persona is a fictional but realistic description of a target user~\cite{cooper2004inmates}, traditionally static and used in both visualization and UX design~\cite{crisan2021baton, van2024understanding, scully2026same}. LLM-based personas can instead be engaged conversationally~\cite{brown2020languagemodelsfewshotlearners, kaate2025youalwaysgetananswer, park2023generativeagents, kocaballi2026workshop} and improve when grounded in interview data~\cite{park2026llmagentsgroundedselfreports}. However, they track prompt phrasing more than stable traits and tend to agree~\cite{shu2024you, sharma2024towards}, and they fall short where domain depth is needed~\cite{satkunarajan2026simulate}.
 
Van den Brandt et al.~\cite{van2024understanding} interviewed twenty genomics visualization authors and derived five personas of genomics experts differing in skill, work orientation, automation, and audience. Coding in ATLAS.ti~\cite{atlasti2024} produced a corpus of excerpts attributable to individual participants and usable as evidence. PersonaCite~\cite{truss2026personacite} supplies the mechanism by which excerpts are retrieved and filtered per question, generating first-person answers with claims attributed to participants for verification. The persona abstains when no relevant evidence meets the retrieval threshold. Hosseini et al.~\cite{hosseini2025llms} also derive personas from interviews but abstract them into profiles supplied up front, whereas ours retrieve coded excerpts at query time and must cite them. Prior work studies persona-conditioned behavior in isolation~\cite{satkunarajan2026simulate} or offers grounded personas as an interviewable artifact~\cite{park2026llmagentsgroundedselfreports}; we embed such personas inside a running evaluation protocol against a live system.

Genomics exemplifies specialized domains. Its tools include GUI systems~\cite{thorvaldsdottir2013integrative}, visualization grammars~\cite{lyi2021gosling}, and retrieval systems~\cite{PandeyGenoREC, nguyen2026geranium}, serving diverse users~\cite{crisan2021baton, welch2014bioinformatics, lyi2023role, walters2025gqvis} who can be hard to fully represent in a study. A Gosling~\cite{lyi2021gosling} JSON specification defines data sources, coordinates, marks, and channels, serving as an adaptable template for different datasets and visualization needs.

\begin{figure*}[!t]
  \centering
  \begin{tikzpicture}[
  font=\sffamily,
  >=Stealth,
  line cap=round,
  line join=round
]

\definecolor{primaryMainBlock}{RGB}{0,92,132}
\definecolor{lightMainBlock}{HTML}{DAE9F7}
\definecolor{mainBg}{HTML}{FAF7F2}

\definecolor{primarySecondaryBlock}{HTML}{9E9E9E}
\definecolor{externalSubBlock}{HTML}{f4f4f4}
\definecolor{externalBg}{HTML}{ffffff}
\definecolor{externalArrow}{HTML}{555555}

\tikzset{
boxMain/.style={
    draw=primaryMainBlock, fill=lightMainBlock, rounded corners=3pt,
    align=center, inner sep=7pt, line width=0.9pt
  },
boxMainInner/.style={
    draw=primaryMainBlock, fill=white, rounded corners=3pt,
    align=center, inner sep=7pt, line width=0.9pt
  },
boxExt/.style={
    draw=gray!55, fill=externalBg, rounded corners=4pt,
    align=center, inner sep=8pt, line width=0.9pt
  },
boxGreen/.style={
    draw=primarySecondaryBlock, fill=externalSubBlock, rounded corners=6pt,
    align=center, inner sep=9pt, line width=0.9pt
  },
groupSyc/.style={
    draw=gray!55, fill=mainBg, rounded corners=10pt,
    inner sep=12pt, line width=0.9pt
  },
groupGer/.style={
    draw=gray!55, fill=externalBg, rounded corners=10pt,
    inner sep=12pt, line width=0.9pt
  },
ragGroup/.style={
    draw=primaryMainBlock, dashed, rounded corners=8pt,
    inner sep=12pt, line width=0.9pt, fill=lightMainBlock!25
  },
primaryFlow/.style={
    -{Triangle[length=2.2mm,width=2mm,fill=primaryMainBlock]},
    draw=primaryMainBlock,
    line width=1.3pt
},
replyFlow/.style={
    -{Straight Barb[length=1.7mm,width=1.9mm]},   %
    draw=primaryMainBlock, line width=1.3pt
},
extFlow/.style={
    -{Latex[length=2mm,width=1.6mm]},
    draw=externalArrow,
    line width=0.8pt,
     dotted
},
stepnode/.style={
    circle, draw=primaryMainBlock, fill=primaryMainBlock,
    text=white, font=\bfseries\scriptsize, inner sep=1.6pt
  },
lab/.style={font=\sffamily\footnotesize, inner sep=1.5pt},
pageStyle/.style={
    draw=gray!35, fill=white, line width=0.8pt, inner sep=16pt
  }
}

\node[boxMain, minimum width=3.4cm] (cli)    at (-2.3, 3.6) {\textbf{Batch interface}\\[2pt]Headless CLI};
\node[boxMain, minimum width=3.4cm] (viewer) at ( 1.9, 3.6) {\textbf{Live session viewer}\\[2pt]Streaming};

\node[boxMain, minimum width=6cm, minimum height=1.2cm] (prot) at (0, 1.18) {\textbf{Protocol runner}};
\node[boxMain, minimum width=2.5cm] (eval) at (-5.8, 1.2) {\textbf{Evaluation records}\\[2pt]JSON transcripts};

\node[boxMain, minimum width=6.9cm, minimum height=2cm,
      below=1.7cm of prot] (syn) {};
\node[font=\sffamily\bfseries] at ([yshift=-4mm]syn.north) {Synthetic evaluators};

\node[boxMainInner, minimum width=2.8cm] (ung)
  at ([xshift=-1.7cm,yshift=-2mm]syn.center) {Ungrounded};

\node[boxMainInner, minimum width=2.8cm] (grd)
  at ([xshift= 1.7cm,yshift=-2mm]syn.center) {Grounded};

\node[boxMain, minimum width=2.4cm, minimum height=1.25cm,
      below=1.55cm of grd] (idx) {\textbf{Evidence index}\\[2pt]Coded excerpts};

\node[boxMain, minimum width=2.8cm, minimum height=1.25cm,
      left=2.5cm of idx] (emb) {};
\node[font=\sffamily, align=center] at (emb.center) {\textbf{Encoder}\\[2pt]MiniLM};
\mascot{($(emb.south east)+(-0.45,0.33)$)}{externalSubBlock!100}{0.8}

\node[boxExt, minimum width=3.4cm, minimum height=1.4cm,
      left=3.3cm of emb] (voc) {\textbf{User-study corpus}\\[2pt]{Interview Study~\cite{van2024understanding}}};
\node[font=\sffamily] at ([yshift=3mm]voc.north) {\textless \code{external}\textgreater};

\node[boxExt, minimum width=3.2cm, minimum height=2cm,
      left=5cm of syn] (vlm) {};
\node[font=\sffamily, align=center]
  at ([yshift=4mm]vlm.center)
  {\shortstack{\textbf{Generative model}\\[2pt]{Vision-Language}}};
\mascot{($(vlm.center)+(0,-.55)$)}{externalSubBlock}{1.2}
\node[font=\sffamily] at ([yshift=3mm]vlm.north) {\textless \code{external}\textgreater};

\node[boxExt, minimum width=3.7cm, minimum height=1.7cm,
    right=3.3cm of prot] (ger) {};
\node[font=\sffamily] at ([yshift=3mm]ger.north) {\textless \code{external}\textgreater};

\draw[primaryFlow] (cli.south)    -- ([xshift=-2.3cm]prot.north);
\node[stepnode] (s1a) at ($(cli.south)!0.4!([xshift=-2.3cm]prot.north)$) {\code{1}};
\node[lab, right=1pt] at (s1a.east) {\code{run()}};

\draw[primaryFlow] (viewer.south) -- ([xshift=1.9cm]prot.north);
\node[stepnode] (s1b) at ($(viewer.south)!0.4!([xshift=1.9cm]prot.north)$) {\code{1}};
\node[lab, right=1pt] at (s1b.east) {\code{run\_streaming()}};

\node[font=\sffamily\bfseries\small] at (-.2,2.45) {OR};

\draw[primaryFlow] (prot.west) -- (eval.east)
  node[pos=0.42,stepnode] {\code{5}}
  node[pos=0.45,lab,above=7pt] {\code{emit()}};

\coordinate (p2down) at ([xshift=-2.3cm]prot.south);  %
\coordinate (p2up)   at ([xshift=-1.75cm]prot.south);  %

\coordinate (p2mid)  at ($(prot.south)!0.5!(syn.north)$);
\coordinate (p2xmid) at ($(p2down)!0.5!(p2up)$);

\draw[primaryFlow] (p2down) -- (p2down |- syn.north);
\node[lab, anchor=east] at (p2down |- p2mid) {\code{elicit\_query()}};

\draw[replyFlow] (p2up |- syn.north) -- (p2up);
\node[lab, anchor=west] at (p2up |- p2mid) {\code{[query]}};

\node[stepnode] (s2) at (p2xmid |- p2mid) {\code{2}};

\coordinate (p4down) at ([xshift=1.35cm]prot.south);  %
\coordinate (p4up)   at ([xshift=1.9cm]prot.south);  %

\coordinate (p4mid)   at ($(prot.south)!0.5!(syn.north)$);
\coordinate (p4xmid)  at ($(p4down)!0.5!(p4up)$);

\draw[primaryFlow] (p4down) -- (p4down |- syn.north);
\node[lab, anchor=east] at (p4down |- p4mid) {\code{respond()}};

\draw[replyFlow] (p4up |- syn.north) -- (p4up);
\node[lab, anchor=west] at (p4up |- p4mid) {\code{[reaction]}};

\node[stepnode] (s4) at (p4xmid |- p4mid) {\code{4}};

\begin{pgfonlayer}{groupbg}
\coordinate (ragTopPad) at ([yshift=1mm]idx.north);
\node[ragGroup, draw=none, fill=none, fit=(idx)(emb)] (rag) {};
\coordinate (sycTopPad) at ([yshift=4mm]viewer.north);
\node[groupSyc, fit=(cli)(viewer)(prot)(eval)(syn)(rag)(sycTopPad)] (syc) {};

\node[ragGroup, fit=(idx)(emb)(ragTopPad)] {};
\end{pgfonlayer}

\coordinate (p3up)   at ([yshift= 2.7mm]prot.east);  %
\coordinate (p3down) at ([yshift=-2.7mm]prot.east);  %

\coordinate (p3upGer)   at ([yshift= 2.7mm]ger.west);
\coordinate (p3downGer) at ([yshift=-2.7mm]ger.west);

\coordinate (p3xmid) at ($(p3up)!0.5!(p3upGer)$);
\coordinate (p3ymid) at ($(p3up)!0.5!(p3down)$);

\draw[primaryFlow] (p3up) -- (p3upGer);
\node[lab, anchor=south, yshift=1pt] at (p3xmid |- p3up) {\code{submit\_query()}};

\draw[replyFlow] (p3downGer) -- (p3down);
\node[lab, align=center, anchor=north, yshift=-1pt]
  at (p3xmid |- p3down)
  {\code{[results]}\\\{\code{{text,image,spec\}}}};

\node[stepnode] (s3) at (p3xmid |- p3ymid) {\code{3}};

\draw[extFlow] (vlm.east) -- ([yshift=0mm]syn.west)
  node[midway,sloped,inner sep=0pt,fill=mainBg,lab] (gen)
       {engine};
       
\draw[extFlow] (voc.east) -- (emb.west)
  node[midway,sloped,inner sep=0pt,fill=mainBg,lab] (coded)
       {user evidence};
  
\draw[extFlow] (emb.east) -- (idx.west)
  node[midway,sloped,inner sep=0pt,fill=lightMainBlock!25,lab] (embed)
       {embeddings};
\node[below=1pt of embed,lab,fill=lightMainBlock!25,inner sep=0.5pt]
  {(offline)};

\draw[extFlow] (idx.north) -- (grd.south);
\node[below=15pt of grd,lab,fill=mainBg,inner sep=0.5pt]
  {top-k evidence (per query)};

\coordinate (ragbar) at (0,-3.1);

\node[font=\sffamily, anchor=north west]
  at ([xshift=2pt,yshift=2pt]rag.north west) {\textbf{RAG}};

\node[font=\sffamily] at ([yshift=-10pt]syc.north) {\textbf{Synthetic-evaluation Framework} $|$ Sycamore};

\node[font=\sffamily] at ([yshift=-26pt]ger.north) {\shortstack{\textbf{Visualization system}\\{\textbf{under evaluation}}\\[2pt]{Geranium~\cite{nguyen2026geranium}}}};

\begin{scope}[shift={(6.7,-4)}]
  \def\arlen{0.95}          %
  \def\gap{0.22}            %
  \def\rowsep{0.62}

  \coordinate (r1) at (0, 0);
  \coordinate (r2) at (0, -\rowsep);
  \coordinate (r3) at (0, -2*\rowsep);
  \coordinate (r4) at (0, -3*\rowsep);

  \draw[primaryFlow] (r1) -- ($(r1)+(\arlen,0)$);
  \node[lab, anchor=west] (Lc) at ($(r1)+(\arlen+\gap,0)$)
    {{Protocol call}};

  \draw[replyFlow] (r2) -- ($(r2)+(\arlen,0)$);
  \node[lab, anchor=west] (Lr) at ($(r2)+(\arlen+\gap,0)$)
    {{Protocol reply}};

  \node[stepnode] (Ls) at ($(r3)+(\arlen/2,0)$) {\code{n}};
  \node[lab, anchor=west] (Lp) at ($(r3)+(\arlen+\gap,0)$)
    {{Protocol step}};

  \draw[extFlow] (r4) -- ($(r4)+(\arlen,0)$);
  \node[lab, anchor=west] (Ld) at ($(r4)+(\arlen+\gap,0)$)
    {{Dependency}};

  \node[inner sep=0] (LA) at ($(r1)+(-0.15, 0.06)$) {};
  \node[inner sep=0] (LB) at ($(r4)+(-0.15,-0.06)$) {};

  \begin{pgfonlayer}{groupbg}
    \node[draw=gray!55, fill=white, rounded corners=6pt, line width=0.9pt,
          inner sep=8pt, fit=(LA)(LB)(Lc)(Lr)(Lp)(Ld)(Ls)] (legendbox) {};
  \end{pgfonlayer}
  \node[font=\sffamily\bfseries, anchor=west]
    at ([yshift=11pt]legendbox.north west) {Legend};
\end{scope}

\end{tikzpicture} %
   \caption{
   \textbf{Running a synthetic evaluator to a live system.}
\bluecircle{1} A batch (headless CLI) or streaming (live viewer) interface starts the protocol.
\bluecircle{2} The protocol runner prompts an evaluator and receives a query.
\bluecircle{3} The runner submits that query, as text, image, or specification, to the system under evaluation (here, Geranium~\cite{nguyen2026geranium}) and receives the results (here, \texttt{\{text,image,spec\}} triplets).
\bluecircle{4} The results, including rendered images, are returned to the evaluator to capture its reaction.
\bluecircle{5} The protocol saves the outputs, here as JSON transcripts.
Stages \bluecircle{2}--\bluecircle{4} repeat for each query and modality.
A \bluebox{Grounded} evaluator draws on a retrieval subsystem \ragbox{\textbf{RAG}} that pulls top-\(k\) evidence from an offline user-study corpus (here, Interview Study~\cite{van2024understanding}), the small \bluebox{Ungrounded} reference run bypasses it; \textless \code{external}\textgreater ~components are substitutable.
}
 \label{fig:architecture}
 \vspace{-1.4em}
\end{figure*}

\section{Grounded Synthetic Evaluators}
\label{sec:approach}

In this work, we define a \textit{synthetic persona} as the characterization of a specific user profile being represented. A \textit{synthetic evaluator} is an LLM instance that evaluates a given stimulus under the constraints of that persona, orchestrated by a \textit{protocol runner}.

\subsection{Constructing an Evaluator}
\label{sec:approach:evaluator}

Because the underlying generating engine is a vision-language model, synthetic reactions are conditioned on both textual inputs and rendered chart images. Evaluators in our system operate in two forms that share this underlying engine: a small \textit{ungrounded} reference run and the \textit{grounded} evaluators that form the primary object of our study. The ungrounded reference evaluator is conditioned solely on a generic profile prompt, lacking empirical backing. 

In contrast, a \textit{grounded} evaluator is instantiated from a documented corpus of real users and is assembled at query time using three core components. First, a \textbf{profile} defines the persona's identity, including position type, organization, self-assessed computational skills, work focus, automation level, and primary audience. Second, an \textbf{evidence index}, built offline from a source interview corpus, stores coded behavioral excerpts. Third, an \textbf{evidence gate} dictates whether the evaluator is permitted to respond at each turn. The evaluator answers in the first person and explicitly cites its source participants; however, it may only draw from excerpts retrieved within its designated participant pool that surpass a relevance filter as similarity threshold. If no relevant evidence is found, the gate forces the evaluator to return a fixed non-answer (abstention). The gate applies both to workflow recall and to judgments of returned results. Turns expressing purely subjective preference are exempt from this strict grounding requirement.

\subsection{Running an Evaluator Through a Protocol}
\label{sec:approach:architecture}

Sycamore introduces a protocol runner connecting evaluators to the system under evaluation (Figure~\ref{fig:architecture}). A session follows five stages, with stages \bluecircle{2} through \bluecircle{4} repeated for each query and modality to produce a multi-turn transcript.

Studies run through either a headless CLI for batch processing or a live session viewer for streaming. Both invoke the same runner, which executes an ordered sequence of two turn types. \textit{Scripted} turns elicit free-text responses; \textit{exploration} turns elicit a query, submit it to the system, and return results to the evaluator for a reaction. The runner remains agnostic to evaluator construction as it does not interpret requests or reactions, thus extending Sycamore to a new system requires writing a single adapter. Sessions are recorded in a unified JSON schema containing queries, results, reactions, citations, and abstentions for subsequent coding.

\section{Instantiation: Genomics Visualization Retrieval}
\label{sec:instantiation}

\subsection{System Under Evaluation: Geranium}
\label{sec:inst:object}

We instantiate Sycamore to evaluate Geranium~\cite{nguyen2026geranium}, a multimodal retrieval system built on the Safire similarity framework~\cite{nguyen2025safire} that helps genomics scientists find and adapt interactive visualization templates. Queries in any of the three \textit{modalities} \texttt{\{image, text, specification\}} return multimodal triplets drawn from 3,200 visualizations in 50 categories, where the specification acts as an editable scaffold. For this work, we evaluate the system configuration in Geranium's published user study~\cite{nguyen2026geranium}.

\subsection{Evaluation Protocol}
\label{sec:inst:protocol}

We adapt the evaluation protocol from the published Geranium user study~\cite{nguyen2026geranium}. The original study comprised three phases: workflow description, tool demonstration, and hands-on exploration, followed by a modality preference ranking (1--3 scale) and rationale. In our synthetic protocol, we retain all phases except the demonstration, which is replaced by a textual feature description. 

The exploration phase is driven by a held-out test suite, where each stimulus is a triplet of a chart image, a Gosling specification, and a natural-language (NL) description. For the image and specification modalities, the evaluator issues the held-out artifact directly. For the text modality, the evaluator formulates a high-level query based on the NL description, ensuring the condition mirrors a realistic information need. Ultimately, each transcript comprises 4 scripted turns with 9 exploration reactions (13 in total), and a concluding modality ranking.

\subsection{Grounded Evaluators}
\label{sec:inst:grounded}

The grounded evaluators are instantiated based on the personas and coded interview data from van den Brandt et al.~\cite{van2024understanding} (Section~\ref{sec:related}). To match the sample size of the original Geranium study, we instantiate $N{=}7$ evaluators across four distinct personas: two Computational Biologists (CB1, CB2), two Bioinformaticians (BIF1, BIF2), two Software Engineers (SE1, SE2), and one Biologist (Bio). These personas are differentiated by skill levels and role emphases (e.g., scientist-engineers versus tool-builders); complete profile definitions are provided in the Supplementary Material. Notably, we exclude the fifth persona, Visualization Expert to match Geranium's target audience of genomics researchers adapting charts for data analysis, acknowledging this decision excludes the persona most fluent to encoding and grammar nuances.

\subsubsection{Instantiation Pipeline}
\label{sec:inst:pipeline}

\begin{figure}[!h]
    \centering
    \vspace{-.3em}
    \includegraphics[width=.99\linewidth]{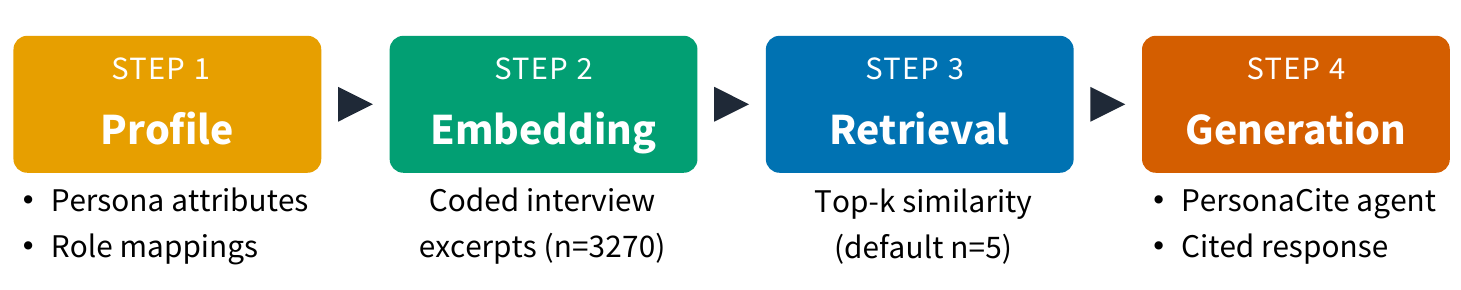}
    \vspace{-.8em}
    \caption{The pipeline for instantiating a grounded evaluator.}
    \label{fig:persona-pipeline}
    \vspace{-.5em}
\end{figure}

We transform the raw persona data into synthetic evaluators using the four-step pipeline illustrated in Figure~\ref{fig:persona-pipeline}. 
\textbf{(1) Profile}: The profile fields described in Section~\ref{sec:approach:evaluator} are populated using the source study, with skills mapped to a 1--3 scale. 
\textbf{(2) Embedding}: A total of 3,270 ATLAS.ti-coded~\cite{atlasti2024} excerpts (across 31 categories) are embedded using \texttt{all-MiniLM-L6-v2}, with each embedding generated from the concatenation of the excerpt's code labels and quote text. 
\textbf{(3) Retrieval}: At query time, the system embeds the incoming question using the same model and retrieves the top-$k$ excerpts ($k{=}5$) from the evaluator's specific participant pool via cosine similarity. 
\textbf{(4) Generation}: As in PersonaCite~\cite{truss2026personacite}, we filter excerpts for relevance, then generate a cited first-person response over the survivors. Filtering uses \texttt{gpt-5-nano}; the harder generation step uses the more capable \texttt{gpt-5-mini}.

\paragraph{Design Rationale} 
Embedding excerpts alongside their code labels allows the qualitative codebook taxonomy to influence semantic similarity, bridging the gap between raw text and the interviewer's analytical decisions. Furthermore, restricting retrieval to an evaluator's distinct participant pool ensures that personas remain separated; otherwise, all seven evaluators would draw from a shared global pool and regress toward a generic mean. A recognized trade-off of this design is that personas mapped to fewer source participants experience higher rates of abstention (Table~\ref{tab:structural}).

\begin{figure}
  \centering
  \vspace{-1em}
  \includegraphics[width=\linewidth]{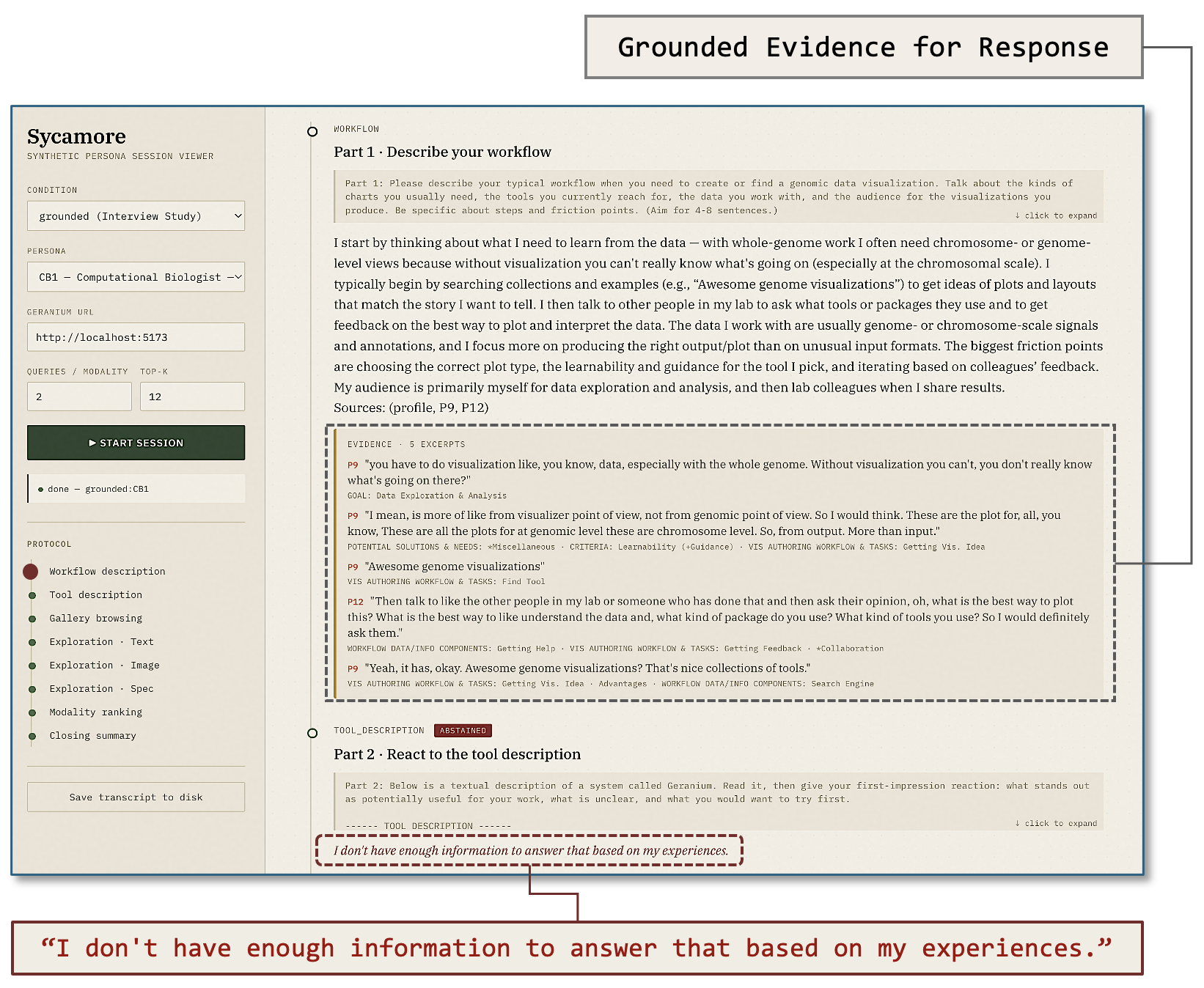}
  \caption{Sycamore viewer streaming a grounded evaluator (Computational Biologist) on Geranium~\cite{nguyen2026geranium}. Responses include cited interview excerpts~\cite{van2024understanding} (e.g., P9), abstaining if no evidence is found.}
  \label{fig:interface}
   \vspace{-1.2em}
 \end{figure}

\paragraph{The Abstention Threshold} 
A critical parameter in the evidence gate is the similarity threshold $\tau$, which we set to $0.45$. This value was derived from the distribution of top-1 similarities produced by protocol questions during a pilot phase. Our choice of threshold is deliberately conservative: during formative evaluation, a missed answer (abstention) incurs minimal cost to the system designer, whereas a fabricated answer lowers trust in the transcript. $\tau$ represents a tunable threshold that controls the abstention rate. Figure~\ref{fig:interface} shows a Computational Biologist evaluator in the live session viewer with cited responses and evidence-based abstention.

\subsection{Two Reference Points}
\label{sec:inst:references}

We contextualize grounded transcripts using two references: (1) an \textbf{ungrounded reference run} with seven evaluators following the same protocol with a generic ``genomics researcher'' prompt, and (2) the published \textbf{Geranium user study} with seven domain experts~\cite{nguyen2026geranium}. The former highlights advantages in citation and abstention behavior, although differences in persona specificity prevent isolating the effect of grounding. The latter provides five themes and individual modality preference scores. We consider this a reference instead of a strict baseline, as the experimental setups are not perfectly equivalent, and the human themes have undergone a layer of qualitative interpretation that the synthetic data has not.

\section{Results and Analysis}
\label{sec:analysis}

\subsection{Coding Procedure}
\label{sec:analysis:coding}

We analyze the 14 resulting transcripts (7 grounded, 7 ungrounded) using a hybrid coding approach. The five themes extracted from the Geranium user study~\cite{nguyen2026geranium} provide our deductive framework, allowing us to evaluate the synthetic transcripts along the same dimensions as the human experts: \textbf{(T1)} usability and usefulness, \textbf{(T2)} modality preference, \textbf{(T3)} variety vs. similarity, \textbf{(T4)} gallery browsing for orientation, and \textbf{(T5)} onboarding and user intent. The codebook and detailed protocol are available in Supplementary Material.

To capture novel behaviors outside these predefined themes, we open-coded two seed transcripts (one grounded, CB1; one ungrounded, U1). This \textbf{\textit{inductive}} pass yielded four content codes: \textbf{(I1)} template adaptation, \textbf{(I2)} data-binding friction, \textbf{(I3)} semantic versus visual matching, and \textbf{(I4)} interactivity and export, alongside two structural codes: \textbf{(I5)} grounding stance and \textbf{(I6)} evidence-boundary abstention. Throughout this section, quotations are labeled with the evaluator's identifier (e.g., BIF1, U1), distinct from the internally cited interview participant numbers (e.g., P6).

\subsection{Outputs from Grounded Evaluators}
\label{sec:analysis:structural}

Table~\ref{tab:structural} summarizes the structural markers extracted from the evaluation records, revealing two primary patterns. First, the grounded evaluators successfully avoided convergence: four evaluators ranked specification first, while three preferred images. Each justified its ranking using persona-specific reasoning anchored in cited evidence. For example, CB1 favored images to \textit{``judge visual fit quickly''} when adapting figures from literature, whereas SE1 preferred specifications due to the ease of swapping data formats (e.g., BigWig, BAM) into a nearly-matched template. Second, every grounded evaluator exhibited evidence-boundary abstention (I6), returning the fixed non-answer on 6 to 9 of their 13 codeable turns (defined in Section~\ref{sec:inst:protocol}) when the retrieval engine failed to find evidence above the similarity threshold. In terms of content, the most prevalent code was template adaptation (I1), where evaluators framed their objective as modifying a near-match rather than authoring from scratch, followed by data-binding challenges (I2).

\begin{table}[!h]
\centering
\small
\vspace{-.5em}
\begin{tabular}{@{}lllcl@{}}
\toprule
Evaluator (Sec.~\ref{sec:inst:grounded}) & Condition & Top Modality & Abstention \# & Cites \\
\midrule
\textbf{BIF1} (Bioinformatician) & grounded & Spec  & 7 & yes \\
\textbf{BIF2} (Bioinformatician) & grounded & Image & 6 & yes \\
\textbf{Bio} (Biologist) & grounded & Image & 6 & yes \\
\textbf{CB1} (Comp. Biologist) & grounded & Image & 6 & yes \\
\textbf{CB2} (Comp. Biologist) & grounded & Spec  & 6 & yes \\
\textbf{SE1} (Software Engineer) & grounded & Spec  & 9 & yes \\
\textbf{SE2} (Software Engineer) & grounded & Spec  & 8 & yes \\
\textbf{U1--U7} & ungrounded & Spec (all) & 0 & no \\
\bottomrule
\end{tabular}
\vspace{1em}
\caption{Characterizations of synthetic evaluators. Top Modality is based on the 1--3 preference ranking (3 = most preferred).}
\label{tab:structural}
\vspace{-1em}
\end{table}

The ungrounded reference run highlights the behaviors introduced by grounding. All seven ungrounded evaluators returned an identical modality ranking (Specification, Image, Text) backed by nearly interchangeable rationales. Furthermore, they exhibited an inflated technical fluency, confidently naming specific embeddings, bin sizes, and genome builds (I5). However, this ungrounded confidence is not falsifiable, whereas claims made by grounded evaluators are directly traceable to source data (e.g., \textit{``something I value when I make interactive genome plots (P6)''}, BIF1). As the grounded evaluators abstained on roughly half of their turns, their final reaction set represents a rigorously \emph{filtered} subset, in contrast to the ungrounded run, which retained a 1:1 volume.

\subsection{Comparison to the Expert User Study}
\label{sec:analysis:vsexpert}

When comparing the synthetic results to the expert reference, modality preference (T2) reveals the most significant divergence. The original human experts generally preferred image-based searches, favoring specifications only when they had fluency with Gosling grammar, and valuing text primarily for quick starts despite inherent ambiguities~\cite{nguyen2026geranium}. The ungrounded run, however, selected specification as the top modality. This likely reflects a failure mode where the LLM over-weights the editable, programmatic logic of the system, mirroring the system's own structural framing~\cite{jain2026sycophancy}. 

In contrast, the grounded transcripts align closer to human expert preferences, with images emerging as a co-leading choice. Their rationales echoed the experts' desire to adapt static figures from the literature, although a slight bias toward specifications remained. Notably, both synthetic conditions under-represented the theme of onboarding (T5). The human experts noted that declarative grammars could feel \textit{``intimidating''} to novices and emphasized that text descriptions support \emph{comprehension} than just querying. This nuance is largely absent from the synthetic runs, plausibly because the source personas (drawn from biomedical informatics) inherently skew toward high technical proficiency. Despite this, both synthetic conditions successfully captured the human tendencies for gallery browsing (T4) and adapting near-matches (I1).

\section{Discussion}

\subsection{The Impact of Grounding on Evaluator Behavior}
Our evaluation indicates that grounding does not necessarily make synthetic evaluators more capable, but rather makes them significantly better calibrated and more diverse. We interpret the uniformity from \textit{ungrounded} instances as largely driven by sycophancy~\cite{jain2026sycophancy}. Grounding introduced persona-level variance and shifted thematic profiles closer to the human reference. This approach trades broad fluency for fewer, attributable responses grounded in empirical quotations, using abstention to avoid fabricated answers. In highly specialized domains such as genomics visualization, this calibration is preferable to the overconfident behavior exhibited by ungrounded models.

\subsection{Practical Considerations for Visualization Studies}

From this case study, we offer provisional considerations for using synthetic evaluators, potentially beyond genomics visualization.

\subsubsection{Select Interview Material \& Build Synthetic Evaluators}

Before constructing evaluators, researchers could consider mapping intended user roles, expertise levels, and evaluation tasks to source participants and relevant interview excerpts. If direct interview material is unavailable, researchers could use published qualitative data with attributable excerpts, narrow the study scope, or conduct focused formative interviews. These interviews could be designed to support later synthetic persona development by including participants with varied roles and expertise levels and documenting concrete workflows and provenance. Ungrounded personas should serve only to generate questions for human investigation, without claims that they cover the intended users.

\subsubsection{Use Synthetic Evaluators in Formative Evaluation}
In the Nested Model~\cite{nested}, abstraction-level validation typically occurs after implementation and depends on access to target users. As recruiting experts is challenging and their time is best reserved for complex analytical tasks, we position grounded synthetic evaluators as an intermediate formative pass at this level: this pass filters trivial usability issues and provides preliminary coverage for personas that might be missing from an upcoming human study (e.g., our synthetic evaluation included software engineers and a biologist, but neither role was heavily represented in the original Geranium study). However, synthetic evaluators cannot replace human judgment, especially regarding trust or long-term tool adoption.

\subsubsection{Prioritize Synthesized Findings for Human Validation}
We recommend distinguishing findings \textit{synthesized} by the model from requirements directly supported by interview excerpts. For example, SE1, SE2, BIF2, and Bio proposed integration with genome browsers such as UCSC and IGV through drill-down links, track hubs, and screenshot queries. Source participants described UCSC-anchored workflows and ease of installation, but none explicitly articulated this requirement. Although the synthesis aligns with recommendations on visualization interoperability~\cite{nguyen2026visualization}, this agreement does not establish a user need. We recommend treating such findings as \textit{testable hypotheses} for targeted follow-up studies, including asking intended users whether and to what extent the proposed integration supports their workflows.

\subsection{Caveats and Limitations}
Synthetic evaluators exhibit biases toward flattery and cannot replicate natural human interaction markers such as hesitation, exploratory clicks, or workflow backtracking. Grounding mitigates but does not eliminate these risks: excerpts reflect the interviewer's framing alongside participants' insights. Synthetic feedback therefore inherits source-data blind spots, requiring careful corpus selection to cover intended personas. Findings for personas absent from the original human study cannot be directly validated here. Finally, Sycamore was instantiated in a single domain with limited runs per persona, leaving variability to be characterized in future work.

\section{Conclusion}

We introduced Sycamore, an exploratory approach to evaluating visualization retrieval with grounded synthetic personas. Its grounding mechanism supports traceable feedback and evidence-based abstention, suggesting a role for synthetic evaluators in formative assessment before human evaluation. Future work will further examine the role of grounding from persona detail and abstention, assess feedback accuracy and calibration, test broader applicability, and validate emergent themes with domain experts.

\acknowledgments{This work was supported in part by the National Institutes of Health (R01HG011773) and by the National Nuclear Security Administration (NNSA) Advanced Simulation and Computing (ASC) Program under grant number DE-SCL0000089.} 

\bibliographystyle{abbrv-doi}
\bibliography{references}
\end{document}